# Data Mining and Electronic Health Records: Selecting Optimal Clinical Treatments in Practice

Casey Bennett, M.A. and Thomas W. Doub, Ph.D.

*Abstract*—Electronic health records (EHR's) are only a first step in capturing and utilizing health-related data - the problem is turning that data into useful information. Models produced via data mining and predictive analysis profile inherited risks and environmental/behavioral factors associated with patient disorders, which can be utilized to generate predictions about treatment outcomes. This can form the backbone of clinical decision support systems driven by live data based on the actual population. The advantage of such an approach based on the actual population is that it is "adaptive". Here, we evaluate the predictive capacity of a clinical EHR of a large mental healthcare provider (~75,000 distinct clients a year) to provide decision support information in a real-world clinical setting. Initial research has achieved a 70% success rate in predicting treatment outcomes using these methods.

## I. INTRODUCTION

RECENT years have the seen the proliferation of electronic health records (EHR's) across the mental healthcare field and the healthcare industry in general. The current challenge is turning data collected within EHRs into useful information. An EHR is only the first step – data must be leveraged through technology to inform clinical practice and decision-making. Without additional technology, EHR's are essentially just copies of paper-based records stored in electronic form.

Centerstone, the largest community-based mental health provider in the United States, is conducting research and development on a number of real-time decision support systems – including areas such as clinical productivity and optimal treatment selection – that combine elements of data mining and predictive modeling with actual clinical practice. Data mining can capture complex patterns about patients' genetic, clinical, and socio-demographic characteristics, which can be used to generate predictions about treatment outcomes. As a result, inherited risks and environmental/behavioral factors associated with patient disorders can be profiled and used to construct the backbone of clinical decision support systems driven by live data based on the actual population.

Beyond this, there is stark evidence of a 13-17 year gap between research and practice in clinical care [1]. This reality suggests that the current methods for moving scientific results into actual clinical care are lacking. Furthermore, evidence-based treatments derived from such research are often out-of-date by the time they reach widespread use and don't always account for real-world variation that typically impedes effective implementation [2]. Indeed, these issues have been a major reason for the push for clinical decision support in healthcare. However, many of the current decision support systems rely on expert-based or standards-based models, rather than data-driven ones. The former are based on statistical averages or expert opinions of what works for groups of people in general, whereas data-driven models are essentially an individualized form of practice-based evidence drawn from the live population. The latter falls within the concept of "personalized medicine."

The ability to adapt specific treatments to fit the characteristics of an individual's disorder transcends the traditional disease model. Prior work in this area has primarily addressed the utility of genetic data to inform individualized care. However, it is likely that the next decade will see the integration of multiple sources of data - genetic, clinical, socio-demographic – to build a more complete profile of the individual, their inherited risks, and the environmental/behavioral factors associated with disorder and the effective treatment thereof [3]. Indeed, we already see the trend of combining clinical and genetic indicators in prediction of cancer prognosis as a way of developing cheaper, more effective prognostic tools [4], [5], [6].

However, data mining alone – or clinical decision support

Casey Bennett is with the Department of Informatics at Centerstone Research Institute, 44 Vantage Way Suite 280 Nashville, TN 37208, (615)460-4111 (email: Casey.Bennett@CenterstoneResearch.org)

Tom Doub is the Chief Operating Officer of Centerstone Research Institute, 44 Vantage Way Suite 280 Nashville, TN 37208, (615)463-6638 (email: Tom.Doub@CenterstoneResearch.org)

Funding for this work was provided by the Ayers Foundation and the Joe C. Davis Foundation.

alone – are only components of a larger potential system. Utilizing them in conjunction creates a system of real-time data-driven clinical decision support, or "adaptive decision support." The result is a more responsive and relevant model, essentially representing a form of rudimentary artificial intelligence that "lives" within the clinical system, can "learn" over time, and can adapt to the variation seen in the actual real-world population. The approach is two-pronged – both developing new knowledge about effective clinical practices as well as modifying existing knowledge and evidence-based models to fit real-world settings.

Continuous improvement of clinical decision support and advancement of clinical knowledge are seen as key features for future systems [7]. In terms of actual application, modeling can be used to support clinical decisions provided a flexible, adaptable IT framework can consolidate data from different sources. Typically, data warehousing provides such an infrastructure. As opposed to the EHR, a data warehouse does not have to be tied to a single provider organization, increasing its power, scope and utility. Patterns learned from past experience can then be applied to new clients as they enter the system (Fig. 1).

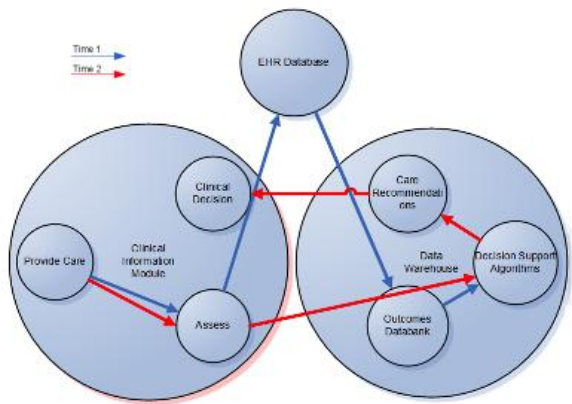

Fig. 1. Clinical Decision Support – Data Flow Diagram

In this study, we describe data mining and predictive modeling work that utilizes clinical indicators to predict client outcomes within the Centerstone system. Those algorithms can then be applied to new clients to aid in selection of the optimal clinical treatment based on a number of possible "service packages". This approach was cross-diagnostic, and the services distinguished only at a gross level (therapy, medical, case management). Nonetheless, results indicate the approach as a promising avenue of research. The initial work was necessitated by changes to a state-run payor (non-Medicaid "Safety Net") in the state of Tennessee, which compelled Centerstone to optimize certain services for the patients most in need. The goal was thus to determine the probability that a given set of services would result in average or above-average outcomes for a particular client and service package. This method would provide clients with the best probability of positive outcomes while minimizing use of services unlikely to result in a positive outcome, increasing the availability of limited resources for other clients.

## II. METHODS

### A. Data Extraction

Data was extracted from Centerstone's electronic health record into a specialized schema in the data warehouse for data mining applications. The target variable was the follow-up CARLA outcome measure (Centerstone Assessment of Recovery Level – Adult) at 6 months post baseline. The CARLA is a measure of level of recovery developed and validated by clinical experts at Centerstone. Using the CARLA, clinicians provide a systematic rating of client symptoms, functioning, supports, insight, and engagement in treatment. Predictor variables initially extracted for the analysis included Baseline CARLA Score, Gender, Race, Age, Baseline Tennessee Outcomes Measurement System (TOMS) Symptomatology Score, Baseline TOMS Functioning Score, Previous Mobile Crisis Encounter (binary, yes/no), Diagnosis Category, Payor, Location, County , Region Type (Urban or Rural), Service Profile (types of services received) and Service Volume (amount of services received). The initial sample was delimited to June 1, 2008 through approximately June 1, 2009 and included only new intakes at time of baseline CARLA (had not seen previously in Centerstone's clinics since at least 2001). After these various filters were applied and data was screened for missing key fields (such as the CARLA at both baseline and follow-up), the final sample size for initial modeling was 423.

### B. Data Modeling

After the initial data extraction and calculations were made, data was loaded into KNIME (Version 2.1.1) [9], an advanced data mining, modeling, and statistical platform. The initial analysis focused on the change in CARLA scores over time. The primary question was whether clients would obtain average or better outcomes based on services received (or vice versa, worse outcomes). As such, the target variable – improvement in clinical outcome – was discretized into a binary variable of plus/minus the mean (equivalent to equal bins classification). The consequences and assumptions of reduction to a binary classification problem are addressed in [6], noting that the issues of making such assumptions are roughly equivalent to making such assumptions around normal distributions. All predictor variables were z-score normalized. Subsequently, all predictor variables were either 1) not discretized (labeled "Bin Target"), or 2) discretized via CAIM (Class-Attribute Interdependence Maximization). CAIM is a form of entropy-based discretization that attempts to maximize the available "information" in the dataset by delineating categories in the predictor variables that relate to classes of the target variable. By identifying and using patterns in the data itself, CAIM has been shown to improve classifier performance [8]. It should be noted that not all models are capable of handling both discretized and continuous variables, and thus both methods were not applied to all modeling methods. Additionally, some methods, such as certain kinds of neural networks or decision trees, may dynamically convert numeric variables into binary or categorical variables as part of their modeling process. As such, even when no pre-discretization was performed, it may have occurred within the modeling process itself.

Multiple models were constructed on the dataset to determine optimal performance using both native, built-in KNIME models as well as models incorporated from WEKA (Waikato Environment for Knowledge Analysis; Version 3.5.6) [10]. Models were generally run using default parameters, though some experimentation was performed to optimize parameters. Namely, in the results shown here, there were three exceptions: 1) Bayesian Network – K2 was set MaxParents=3, 2) MP Neural Network was set Decay=True, and 3) K-Nearest Neighbors as set KNN=5. Models tested included Naïve Bayes[10], HNB (Hidden Naïve Bayes [11]), AODE (Aggregating One-Dependence Estimators [12]), Bayesian Networks [10], Multi-layer Perceptron neural networks [10], Random Forests [13], J48 Decision Trees (a variant of the classic C4.5 algorithm [14]), Log Regression, and K-Nearest Neighbors [15], Additionally, ensembles were built using a combination of Naïve Bayes, Multi-layer Perceptron neural network, Random Forests, K-nearest neighbors, and logistic regression, employing forward selection optimized by AUC [16]. Voting by committee was also performed with those same five methods, based on maximum probability [17]. Due to the number of models used, detailed explanations of individual methods are not provided here for brevity, but can be found elsewhere (e.g. [10] and [18], and references therein).

*C. Model Evaluation*

Model performance was determined using 10-fold cross-validation [10]. All models were evaluated using multiple performance metrics, including raw predictive accuracy; variables related to standard ROC (receiver operating characteristic) analysis, the AUC (area under the curve), the true positive rate, and the false positive rate [19] and Hand's *H* [20]. The data mining methodology and reporting is in keeping with recommended guidelines [21], [22], such as the proper construction of cross-validation, incorporation of feature selection within cross-validation folds, testing of multiple methods, and reporting of multiple metrics of performance, among others.

Additionally, some of the better performing models were evaluated using feature selection prior to modeling (but within each cross-validation fold). The feature selection methods used include univariate filter methods (Chi-squared, Relief-F), multivariate subset methods (Consistency-Based – Best First Search, Symmetrical Uncertainty Correlation-Based Subset Evaluator) and wrapper-based (Rank Search employing Chi-squared and Gain Ratio). The advantages and disadvantages of these different types of feature selection are well-addressed elsewhere [23].

### III. RESULTS

The results of the various combinations of modeling method and discretization can be seen in Table 1, sorted by AUC. The highest accuracies are between 70-72%, with AUC values ranging .75-.79. It should also be noted that the Spearman's rank-order correlation between AUC and Hand's H was .977 (p<.01), indicating little divergence between the two measures, at least on this particular dataset. Hand [20] has indicated that these two measures will diverge when misclassification costs vary across methods. We found no evidence of that in this case, or at least that the issue was not significant. These initial results suggest a predictive capacity of the current EHR data within Centerstone. We suspect that utilizing outcome measures designed to specifically measure change over time will improve this capacity.

Table 1: Model Performance

| Model | Binning | Accuracy | AUC | TP rate | FP rate | H |
|---|---|---|---|---|---|---|
| \multicolumn{7}{c}{10X Cross-Val (partitioned)} | | | | | | |
| AODE | CAIM | 72.3% | 0.7769 | 74.6% | 32.6% | 0.2739 |
| Lazy Bayesian Rules | CAIM | 71.2% | 0.7741 | 75.2% | 36.2% | 0.2695 |
| Naïve Bayes | CAIM | 71.6% | 0.7706 | 76.5% | 36.5% | 0.2705 |
| Bayes Net - K2 | CAIM | 70.7% | 0.7690 | 75.4% | 37.4% | 0.2550 |
| Bayes Net - K2 | Bin Target | 70.4% | 0.7677 | 75.7% | 38.1% | 0.2561 |
| Ensemble | CAIM | 70.9% | 0.7604 | 76.9% | 38.1% | 0.2452 |
| Naïve Bayes | Bin Target | 68.6% | 0.7587 | 74.7% | 41.0% | 0.2410 |
| Bayes Net - TAN | CAIM | 70.0% | 0.7567 | 73.3% | 37.0% | 0.2302 |
| Bayes Net - TAN | Bin Target | 69.7% | 0.7561 | 73.4% | 37.6% | 0.2392 |
| MP Neural Net | CAIM | 70.7% | 0.7534 | 75.6% | 37.6% | 0.2273 |
| Ensemble | Bin Target | 70.2% | 0.7500 | 74.5% | 37.6% | 0.2195 |
| Classif via Linear Reg | Bin Target | 68.8% | 0.7486 | 71.5% | 37.6% | 0.2356 |
| MP Neural Net | Bin Target | 69.5% | 0.7467 | 73.0% | 37.7% | 0.2368 |
| K-Nearest Neighbor | CAIM | 69.5% | 0.7377 | 73.6% | 38.4% | 0.2093 |
| Vote | CAIM | 68.1% | 0.7362 | 72.7% | 40.5% | 0.2011 |
| Random Forest | Bin Target | 66.0% | 0.7238 | 70.3% | 43.1% | 0.2040 |
| Random Forest | CAIM | 67.8% | 0.7222 | 71.7% | 40.1% | 0.1896 |
| Log Regression | CAIM | 67.8% | 0.7206 | 77.7% | 47.9% | 0.1812 |
| Log Regression | Bin Target | 67.1% | 0.7117 | 71.7% | 41.7% | 0.1799 |
| J48 Tree | CAIM | 68.1% | 0.6813 | 71.5% | 39.4% | 0.1688 |
| Vote | Bin Target | 63.4% | 0.6609 | 76.2% | 57.1% | 0.1239 |
| J48 Tree | Bin Target | 66.9% | 0.6544 | 72.4% | 32.6% | 0.1487 |
| K-Nearest Neighbor | Bin Target | 63.8% | 0.6359 | 65.9% | 44.2% | 0.0786 |

These models were then applied to a series of pre-determined "service packages" that most common clients receive. Client data is Cartesian-joined to the service packages to produce predictions for combinations of each and each service package, in essence a "what if" analysis. The results of one of the higher performing models by AUC

(Bayesian Network – K2) were used to generate predictive information at the time of intake for the clinician. Implementation with the live system is being addressed in a separate, upcoming study (data not shown). However, examples of these predictions (based on actual data) can be seen in Figures 2 and 3.

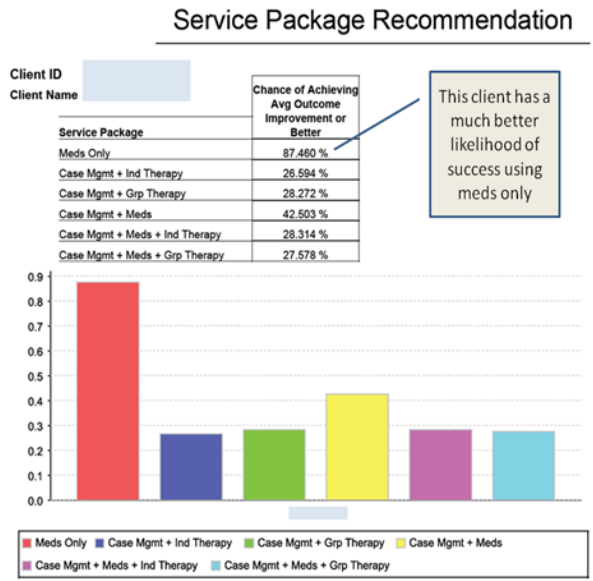

Fig. 2. Example 1 of treatment recommendations using pre-set "service packages"

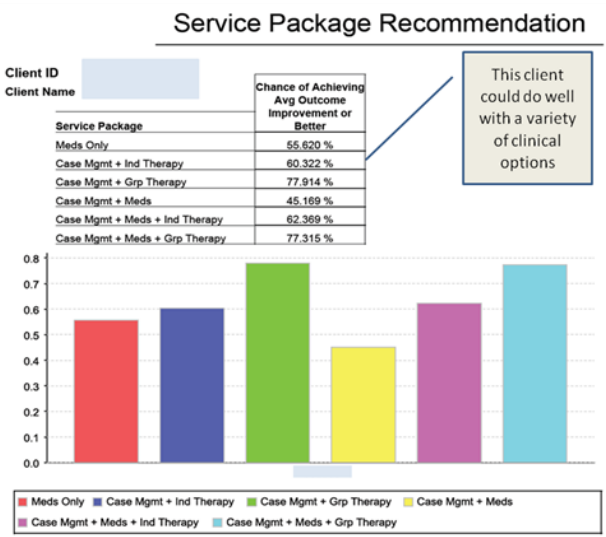

Fig. 3. Example 2 of treatment recommendations using pre-set "service packages"

The results of feature selection were mixed (data not shown). Although some methods were able to produce similar performance using smaller, more parsimonious feature sets than the full feature set models (most notably wrapper-based approaches), they generally did not improve performance significantly. Additionally, the selected feature sets displayed a marked degree of variability across methodologies. This is a common issue, to be expected with complex problems [24], [25]. In many domains there are potentially multiple models/feature sets that can produce comparably good results.

## IV. CONCLUSION

Predictive modeling on live EHR data from a large community-based mental health provider has revealed the capacity of such systems to be used as the basis for an "adaptive" clinical decision support framework. Even without work to enhance the EHR, models were built with over 70% accuracy and .75 AUC in predicting optimal clinical treatments. It is likely through such an approach that the true potential of EHR's can be realized. Indeed, even recent popular media articles are picking up on this distinction between current EHR's and their trumpeted potential ("Little Benefit Seen, So Far, in Electronic Patient Records" New York Times, 11/15/2009).

Without individualized care recommendations that have the capacity to rapidly incorporate changing evidence, adoption of evidence-based practice and treatment guidelines will continue to occur at a glacial pace. While there are many barriers to adoption of systematic treatment recommendations, one of the primary failings of common treatment recommendations is that they are based on statistical averages (e.g., "70% of people improve with medication X"). While those guidelines are helpful, practicing clinicians are keenly aware that treatments that have been shown to be highly effective in clinical trial populations are not always effective with individuals in real-world settings. This is particularly true when the source research failed to adequately address variation due to gender, race, ethnicity, or a multitude of other potential factors. When recommended treatments don't work, clinicians are frequently left to make critical decisions where research evidence is lacking.

The integration of electronic health records with rigorous data modeling as described herein can complement traditional research methods by filling gaps in knowledge, by suggesting new and possibly unanticipated avenues for systematic research, and by enabling rapid deployment of personalized evidence in field settings. New innovations for individualized care can literally be integrated into predictive models overnight, as opposed to the decades that research evidence often takes to diffuse into common practice. This is key to leveraging electronic health data. Without modeling, EHR's are only informative of what occurred in the past, not predictive of the future. Without that predictive capacity, it cannot be used by clinicians as actionable information.

One limitation of this approach is that it requires large and diverse populations, diversity in practice, and reliable data. A small medical practice or group practice could not generate enough data to produce reliable and replicable findings. It is therefore important for provider organizations to consider how to aggregate their data so that predictive models may be developed and fed back into local electronic health records. Privacy and security of health information will have to be paramount, or the risks to individual privacy may outweigh the collective benefit of data aggregation and prevent meaningful advances in care.

It is also worth noting that these models actually benefit from natural variation in clinical practice. The current drive toward standardization and consistency in treatment may actually inhibit innovation that would be identified through modeling efforts. While health research is generally informed by clinical theory and practice, most practice occurs outside the purview of academic medicine, and many exceptional clinical practitioners do not conduct research or publish innovations they may develop with their patients. Modeling can identify emerging clinical practices that are especially promising, and may accelerate the process of dissemination from one clinician to another.

In the work described here, the initial model was built across all diagnoses - including diagnosis used as a predictor variable – but work is proceeding to focus on building models that make personalized clinical predictions within diagnostic-specific groups. Furthermore, research is continuing into analyzing more specific questions, moving from – "does the client need medications" – to "which medications are most likely to be effective for this particular individual". Mixing genetic (e.g. microarray) and clinical indicators, rather than using one or the other, is the most likely long-term avenue, although if and how these data sources should be combined is still a subject of intense debate [4], [5], [6].

The purpose of this study was to test the feasibility of building clinically predictive models using data already existent in the EHR. After this initial work, a new study is currently underway using improved outcome measures that are putatively better indicators of clinical improvement (CDOI – Client-Directed Outcome-Informed, [26]). This is to be followed by the first controlled pilot study on actual implementation of this data-driven decision support model specifically for patients with depression at certain clinical sites. In addition, we are developing a national data warehouse across several major mental healthcare providers from Michigan to Colorado to Arizona, in partnership with the Centerstone Research Institute's Knowledge Network, a technology-based alliance of providers, academic researchers, and industry leaders. Funding is also being sought to develop a gene expression database on a large portion of Centerstone's clinical population, likely starting with clients with depressive disorders or schizophrenia. These future efforts will help to improve and validate these models.

ACKNOWLEDGEMENTS

The authors would like to thank the Ayers Foundation and the Joe C. Davis Foundation for their support in these efforts. The authors would also like to recognize April Bragg, PhD, for her assistance in manuscript preparation.